\documentclass{nature2}

\usepackage{astjnlabbrev-nature}

\usepackage{epsfig}
\usepackage{color}
\usepackage{graphicx}
\usepackage[caption=false]{subfig}
\usepackage{longtable}
\usepackage{bm}
\usepackage{graphicx}
\usepackage{hyperref}

\hypersetup{
    colorlinks=true,
    linkcolor=blue,
    citecolor=blue,      
    urlcolor=blue,
    }

\usepackage{enumitem}
\setlist{nolistsep} 
\setitemize[0]{leftmargin=*}
\usepackage{multirow}
\usepackage{colortbl}
\usepackage{tabulary}
\usepackage{amsmath}
\usepackage{bm}
\usepackage{amssymb}
\usepackage{graphicx}
\usepackage{tikz,lipsum,lmodern}

\usepackage{graphics,graphicx}
\usepackage{hanging}

\usepackage{longtable}
\usepackage[labelsep=endash]{caption}

\title{Gas Phase Metallicities of Local Ultra-Luminous Infrared Galaxies Follow Normal Star-Forming Galaxies}

\author{Nima~Chartab$^{1}$, Asantha~Cooray$^{1}$, Jingzhe~Ma$^{2}$,
  Hooshang~Nayyeri$^{1}$, Preston~Zilliot$^{1}$, Jonathan~Lopez$^{1}$,
  Dario~Fadda$^{3}$, Rodrigo~Herrera-Camus$^{4}$,
  Matthew~Malkan$^{5}$, Dimitra~Rigopoulou$^6$, Kartik~Sheth$^7$,
  Julie Wardlow$^8$}

\makeatletter
\let\saved@includegraphics\includegraphics
\AtBeginDocument{\let\includegraphics\saved@includegraphics}
\makeatother

\begin{document}
\maketitle
\begin{affiliations}
\item {\small Department of Physics \& Astronomy, University of California, Irvine, CA 92697, USA}
\item {\small Center for Astrophysics, Harvard \& Smithsonian, 60 Garden
St, Cambridge, MA 02138, USA}
\item {\small USRA SOFIA, NASA Ames Research Center, MS N232-12, Moffett Field, CA 94035-1000, USA}
  \item {\small Astronomy Department, Universidad de Concepción, Av. Esteban Iturra s/n Barrio Universitario, Casilla 160, Concepción, Chile}
  \item {\small Department of Physics and Astronomy, UCLA, 475 Portola Plaza, Los Angeles, CA 90095, USA}
  \item {\small Astrophysics, Department of Physics, University of Oxford, Keble Road, Oxford OX1 3RH, UK}
  \item {\small Mary W Jackson NASA Headquarters, 300 E Street SW, Washington
DC 20546, USA}
    \item {\small Department of Physics, Lancaster University, Lancaster, LA1
4YB, UK}

\end{affiliations}

\begin{abstract}

  Despite advances in observational data, theoretical models, and computational techniques to simulate key physical processes in the formation and evolution of galaxies, the stellar mass assembly of galaxies still remains an unsolved problem today. Optical spectroscopic measurements appear to show that the gas-phase metallicities of local ultra-luminous infrared galaxies (ULIRGs)
  are significantly lower than those of normal star-forming galaxies\cite{Liang04,Rupke08,Roseboom12}. This difference has resulted in the claim that ULIRGs are fueled by metal-poor gas accretion from the outskirts\cite{Mannucci10}. Here we report on a new set of gas-phase metallicity measurements making use of the far-infrared spectral lines of [O{\sc iii}]52 $\mu$m, [O{\sc iii}]88 $\mu$m, and [N{\sc iii}]57 $\mu$m instead of the usual optical lines. Photoionization models have resulted in a metallicity diagnostic based on these three lines that break the electron density degeneracy and reduce the scatter of the correlation significantly\cite{Pereira-Santaella17}. Using new data from SOFIA and archival data from Herschel Space Observatory, we find that local ULIRGs lie on the mass-metallicity relation of star-forming galaxies and have metallicities comparable to other galaxies with similar stellar masses and star formation rates. The lack of a departure suggests that ULIRGs follow the same mass assembly mechanism as luminous star-forming galaxies and $\sim 0.3$ dex under-abundance in metallicities derived from optical lines is a result of heavily obscured metal-rich gas which has a negligible effect when using the FIR line diagnostics.

\end{abstract}

Galaxy mass assembly is intricately connected to how galaxies form and grow their metal content. Elemental abundances of galaxies, as determined through nebular recombination lines from the HII regions of
the interstellar medium, are used to capture insights into the history of star formation, stellar nucleosynthesis, and baryon recycling processes within galaxies\cite{Maiolino19}.
The mass assembly of galaxies, and thus the metal abundance, is regulated by a diverse set of phenomena, from those associated with the external environment\cite{Chartab21,Sattari21} with gas inflow and mergers, to internal processes such as feedback from the central super-massive black hole appearing as an active galactic nucleus (AGN). Accurate measurement of elemental abundances of different galaxy populations throughout cosmic history remains one of the key observational goals of modern-day astrophysics. 

Ultra-luminous infrared galaxies (ULIRGs) are a cosmologically important galaxy population whose nature changes substantially with redshift. At $z < 0.3$, ULIRGs are rare, with less than one per $\sim$ hundred square degrees, and are invariably mergers between approximately equal-mass galaxies\cite{Clements96,Farrah01,Veilleux02,Veilleux06}. Evidence suggests that the infrared emission that dominates the spectral energy distribution (SED) arises from the dust heated by UV photons, mostly emitted by massive stars associated with the starburst\cite{Genzel98,Franceschini03}, with some role for AGNs\cite{Rigopoulou99,Vega08,Nardini11}. The number of ULIRGs rises rapidly and reaches a density of several hundred per square degree at $z\geq 1$. High-redshift ULIRGs have a lower merger fraction, wider range in dust temperatures and thus SED shapes, and a greater star-formation efficiency compared to local ULIRGs that are dominated by nuclear starbursts\cite{Kartaltepe10,Magdis10,Sajina12,Geach13}. Understanding the physical processes in low-redshift ULIRGs then becomes crucial since they connect to the stellar history and super-massive black hole mass assembly in $\geq L_\star$ galaxies, establishing galaxy evolution over cosmic time for massive galaxies in the Universe.

The gas-phase metal abundances of local ULIRGs, with oxygen abundance used as a proxy for metallicity, inferred from optical emission lines\cite{Caputi08,Rupke08,Kilerci14}
appear to lie below the now well-established stellar mass-metallicity relation for star-forming galaxies, when the two populations are compared at the same stellar mass.
These measurements make use of the same nebular emission lines as used for metallicity measurements of other star-forming galaxies\cite{Kewley08}.
This observed abundance offset from the mass-metallicity relation has at least two explanations\cite{Herrera-Camus18b}. First, as shown by theoretical models and numerical simulations\cite{Montuori10,Rupke10,Torrey12}, tidal forces acting in merging/interacting galaxies primarily funnel low-metallicity gas from the outskirts toward the central active star-forming regions, explaining not only the observed nuclear metallicity under-abundances but also the shallower metallicity gradients\cite{Rupke08,Kewley10,Kilerci14}. If this explanation is correct, then it implies ULIRGs undergo different mass assembly history than those of normal star-forming galaxies. The other explanation is that low gas-phase metal abundances inferred from the optical nebular lines may not be representative of the metallicity of the heavily obscured bulk of the star-forming gas in ULIRGs due to the presence of a large dust mass in ULIRGs associated with the starburst. While the metallicity based on oxygen-to-hydrogen ratio is found to be low, ULIRGs also show a $\sim 3\times$ over-abundance in neon relative to solar\cite{Veilleux09}, suggesting the presence of uncertainties and complexities in measurements and their interpretation. Given the overall implications to understand the mass assembly of galaxies, it becomes necessary to conduct independent, reliable, and extinction-insensitive determinations of gas-phase metallicities to further understand if ULIRGs are, in fact, metal poor relative to other star-forming galaxies.

The far-infrared (FIR) fine-structure lines of nitrogen and oxygen offer a powerful tool to characterize the interstellar medium of local and high-z galaxies, including radiation fields, gas densities, temperatures, and metal abundances that are much less susceptible to extinction than UV and optical transitions\cite{Kaufman06,Fischer14}. 
While UV/optical line diagnostics have been established and used for decades, the use of FIR diagnostics is still in their infancy\cite{Liu01,Nagao11}. As the FIR part of the electromagnetic spectrum is not accessible from the ground, a major limitation associated with FIR diagnostics is the lack of observational facilities with instruments having the necessary sensitivity to conduct sensitive spectroscopic measurements. The Herschel Space Observatory\cite{Pilbratt10} with PACS\cite{Poglitsch10} and SPIRE\cite{Griffin10} instruments offered significantly improved sensitivity and resolution over previous space-based facilities, but observations still lacked the full coverage in the far-infrared band, missing some of the key diagnostic lines for ULIRGs at low redshifts. This gap at wavelengths below 55$\mu$m
is now filled by the Stratospheric Observatory for Infrared Astronomy (SOFIA) with its FIFI-LS instrument providing sensitivity to detect key spectral lines missed by Herschel.

Using SOFIA/FIFI-LS, we observed a sample of five local ULIRGs with $S_{60\mu m} \geq 7$ Jy at $0.01 < z < 0.13$, with key spectral emission lines appearing in windows uncontaminated by the atmospheric emission even at altitude. The observations in the SOFIA/FIFI-LS blue channel cover the [O{\sc iii}]52 $\mu$m and [N{\sc iii}]57 $\mu$m lines missing from archival Herschel/PACS observations (Table \ref{table:line_measurements}; see Methods for measurement details). These spectral lines complete the line observations needed to measure the robust FIR metallicity tracer\cite{Pereira-Santaella17} using the ratio (2.2$\times$[O{\sc iii}]88+[O{\sc iii}]52)/[N{\sc iii}]57. Attempts to estimate abundances with FIR fine-structure emission line diagnostics similar to this ratio using all three spectral lines have been subject to large uncertainties due to the lack of [O{\sc iii}]52 $\mu$m data\cite{Fernandez21}. Therefore, previous attempts to measure metal abundances with FIR emission lines have involved the ratio of [O{\sc iii}]88/[N{\sc iii}]57. Photoionization models, however, show that such a ratio is subject to large uncertainties in the electron density of the ISM and the ionization parameter\cite{Pereira-Santaella17}. Similarly, [O{\sc iii}]52/[N{\sc iii}]57 can be used as a diagnostic to measure gas-phase metal abundance, but again this ratio is impacted by uncertainties in the electron density and ionization parameter. The dependence here, however, is opposite to that of the [O{\sc iii}]88/[N{\sc iii}]57 ratio such that an optimal combination involving both [O{\sc iii}]52 $\mu$m and [O{\sc iii}]88 $\mu$m emission lines is mostly independent of the density. The ratio is capable of providing a metallicity mostly independent of the ionizing source and ionization parameter with an intrinsic scatter of 0.2 dex at a given value of the line ratio\cite{Pereira-Santaella17}. This FIR metallicity estimator is also not significantly affected by either the age of the ionizing stellar population or the presence of an AGN, allowing a clear observational approach to draw definite conclusions on whether the optical-based metallicities are underestimated.
We convert the line ratio, 2.2$\times$[O{\sc iii}]88+[O{\sc iii}]52)/[N{\sc iii}]57, to gas-phase metallicity, adopting the ionization parameter value of $\rm U=10^{-3}$. For our sample with $\rm \Sigma_{FIR}\sim 10^{11}-10^{12}$ $\rm L_\odot/Kpc^2$, an ionization parameter of $\rm U\sim 10^{-2.7}-10^{-3.5}$ is estimated from observations\cite{Herrera-Camus18b}. The existing photoionization models\cite{Pereira-Santaella17} show that this range of ionization parameter has a minimal effect, $<0.05$  dex, in our metallicity measurements.

The metallicity of our ULIRG sample is also measured from optical fine-structure lines using the R23 ratio given by ([O{\sc ii}]3726+[O{\sc iii}]4959, 5007)/H$\beta$ ratio or O3N2 given by ([O{\sc iii}]5007/H$\beta$)/([N{\sc ii}]6584/H$\alpha$), and adopting a theoretical calibration to convert these line ratios using optical nebular recombination lines to an estimate of the oxygen abundance\cite{tre04}. The calibration used to derive optical metallicities\cite{tre04} assumes a different $\rm N/O$ abundance\cite{Charlot01} at a given metallicity compared to the FIR metallicity calibration\cite{Pereira-Santaella17} based on the observed N/O abundances in nearby galaxies\cite{Pilyugin16}. In order to maintain comparable results on the oxygen abundance that can be directly compared to existing results from optical measurements on the mass-metallicity relation of star-forming galaxies, we convert the FIR oxygen abundance to the same metallicity calibration as optical measurements (see Figure \ref{fig:calibration} in the Methods Section). In the Methods Section, we show that our results are also valid even when the original FIR metallicities are retained and compared with optical metallicities derived from a calibration with the consistent assumption of a $\rm N/O-\rm O/H$ relation as FIR metallicity.  

\begin{figure*}
    \centering
    \includegraphics[width=\linewidth]{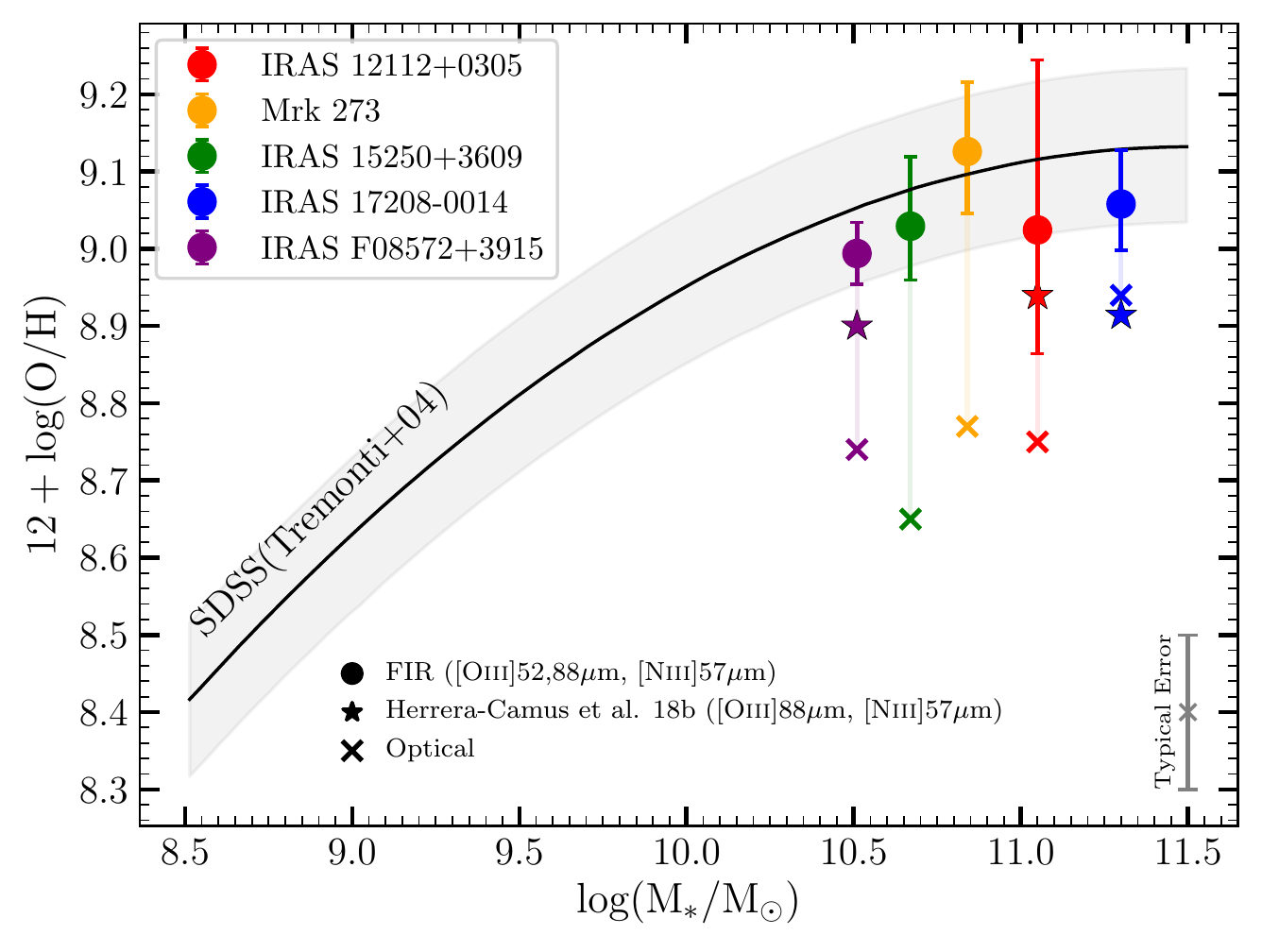}
    \caption{Mass-metallicity relation observed in local galaxies\cite{tre04} and ULIRGs whose metallicities are measured using optical-based (crosses) and IR-based (circles; this work) methods. All metallicity measurements are shown in the same scale\cite{tre04}. The FIR metallicity measurements from literature\cite{Herrera-Camus18b} based on [O{\sc iii}]88/[N{\sc iii}]57 ratio are also included where they are available (stars).}
    \label{fig:MZR}
\end{figure*}

As shown in Figure \ref{fig:MZR}, for the same five ULIRGs, optical recombination line ratio results in metallicity measurements that are at least 0.3 dex lower than the mass-metallicity relation of star-forming galaxies as found with SDSS\cite{tre04}. We also include previously published\cite{Herrera-Camus18b} FIR metal abundance measurements based on [O{\sc iii}]88/[N{\sc iii}]57 ratio where available. They lead to metal abundances that are larger than optical lines estimates but still fall below the mass-metallicity relation of star-forming galaxies\cite{Herrera-Camus18b}. Finally, using [O{\sc iii}]52, 88 $\mu$m and [N{\sc iii}]57 $\mu$m lines, we perform metallicity measurements utilizing the above diagnostic, rescaled to the same oxygen abundance calibration as that used for the mass-metallicity relation with optical line ratios in Figure \ref{fig:MZR}. These FIR-based, extinction-insensitive metallicity measurements indicate that ULIRGs lie on the mass-metallicity relation of star-forming galaxies. They also do not indicate unusual metal deficiencies in ULIRGs, as one would conclude with optical line ratios alone.

\begin{figure*}
    \centering
    \includegraphics[width=\linewidth]{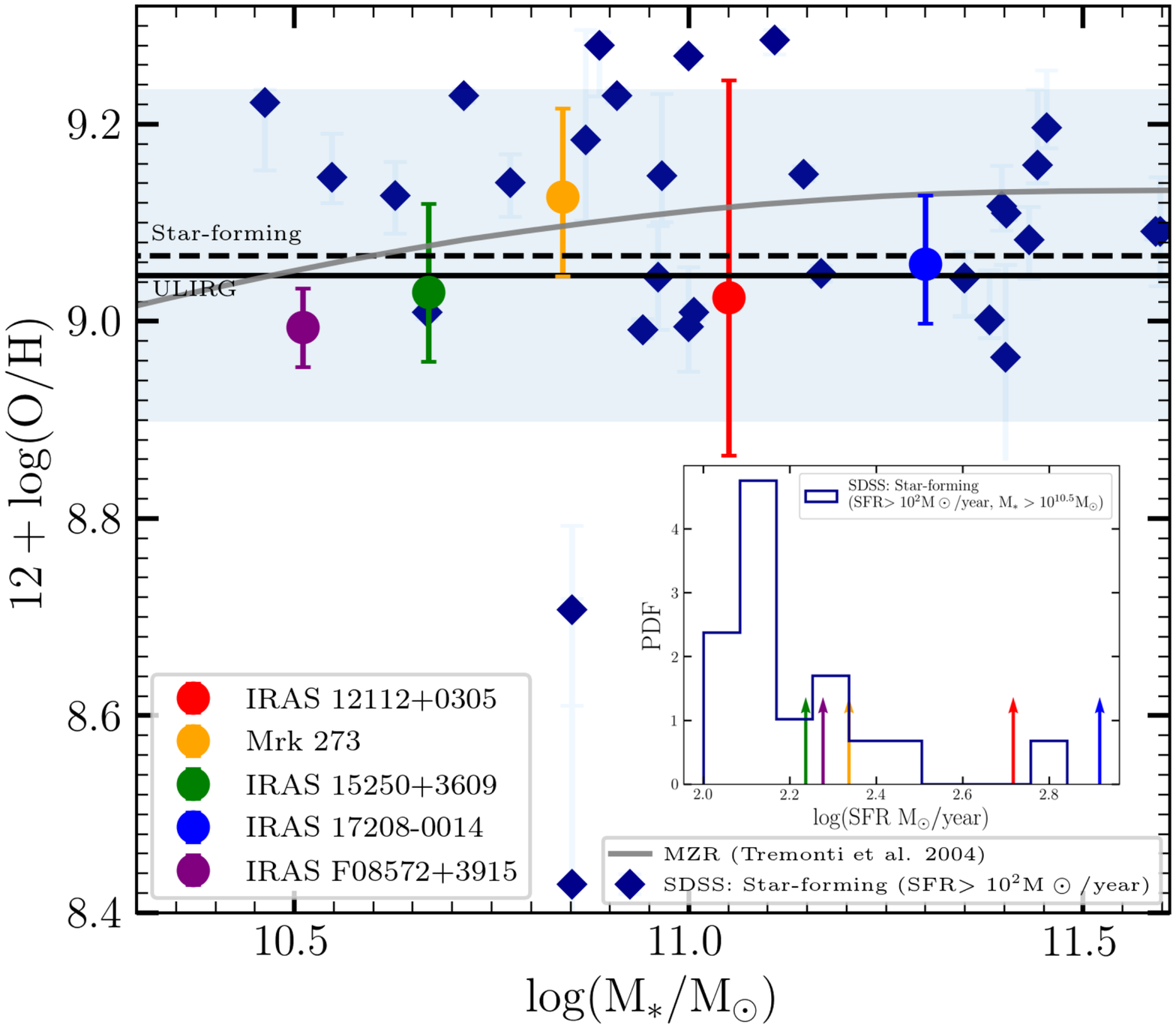}
    \caption{A comparison between the gas-phase metallicity measurements of ULIRGs derived from FIR diagnostic lines and star-forming galaxies with similar star formation rates, excluding extremely dusty galaxies ($\rm A_v<2.4$), derived from optical diagnostic lines. The sub-panel shows the star formation rate distribution of the sample used in this figure. We find that the average FIR-based gas-phase metallicities of ULIRGs (horizontal black line) are consistent with star-forming galaxies (horizontal dashed line). The shaded region shows the 1$\sigma$ scatter in the gas-phase metallicity of the star-forming sample.}
    \label{fig:SDSS}
\end{figure*}

Figure \ref{fig:MZR} also shows that ULIRGs are found primarily towards the bottom of the mass-metallicity relation. We find that this is also consistent with expectations for normal star-forming galaxies with similar star-formation rates as the ULIRG sample. We estimate the average star-formation rate of our ULIRG sample, $\rm SFR\sim 10^{2.5}\ M_\odot/year $, using a calibration between FIR luminosity and star-formation rate\cite{Kennicutt98}. We use the MPA/JHU Value-Added Galaxy Catalog from SDSS-DR7\cite{Kauffmann03,tre04,Brinchmann04} to build a subsample of SDSS galaxies with similar star-formation rates, excluding extremely dusty galaxies. We select galaxies at $0.07<z<0.3$ with $\rm SFR> 10^{2}\ M_\odot/year $ and $\rm A_v<2.4$. We further exclude AGNs from the sample as it is known that optical-based metallicities can be contaminated by the ionizing spectrum of an AGN, while the FIR metallicity indicator used in this work is robust even in the presence of an AGN\cite{Pereira-Santaella17}. These criteria leave us with 35 star-forming galaxies. Figure \ref{fig:SDSS} shows the comparison between the gas-phase metallicity of these galaxies derived from optical diagnostic lines and that of ULIRGs from FIR-based calibration\cite{Pereira-Santaella17}. We find that the gas-phase metallicity of ULIRGs is consistent with star-forming galaxies with similar star-formation rates. ULIRGs do not show a sign of significant metal deficiency.             
According to our result, ULIRGs are metal-rich or at least have oxygen abundances comparable to normal star-forming galaxies but remain heavily dust-obscured. The dust obscuration is an important consideration since the attenuation of optical recombination lines from the HII regions of the ISM could lead to gas-phase metallicity results that are likely to be biased to lower values. The FIR-based metallicity tracer used in this work with all three emission lines has the potential to offer unbiased studies of gas-phase oxygen abundance of the ISM of ULIRGs. While our measurements are for a sample of local ULIRGs, we expect this implication to hold true for all dusty galaxies, including those that are found at high redshifts. Given that the abundance of ULIRG-like dust-rich galaxies increase rapidly at $z > 1$, and even dominating the cosmic star-formation rate density at $z\sim 2-3$, gas-phase metal abundances at the peak epoch of galaxy formation could be impacted at some level if metal abundances are only estimated with rest-frame optical lines. In the future, Origins Space Telescope\cite{Battersby18} has the ability to carry out metal abundance measurements and other diagnostic studies on the ISM and AGN activity in galaxies using a suite of far-infrared emission lines. Further development of metallicity diagnostics such as the ratio used here to properly understand ways to reduce the existing degeneracies will also be useful for the interpretation of results from future infrared facilities.

\newpage
            
\begin{table*}[h]
\centering
\resizebox{1\columnwidth}{!}{%

\begin{tabular}{c c c c} 
 \hline
Source ID & \begin{tabular}{c}[O {\footnotesize III}]52$\mu m$\\ $\times 10^{-16}\rm W.m^{-2}$\end{tabular} & \begin{tabular}{c}[N {\footnotesize III}]57$\mu m$\\ $\times 10^{-16} \rm W.m^{-2}$\end{tabular} & \begin{tabular}{c}[O {\footnotesize III}]88$\mu m$\\ $\times 10^{-16}\rm W.m^{-2}$\end{tabular} \\ [0.5ex] 
 \hline\hline
IRAS 12112+0305 & $3.06\pm 0.69 $ & $0.56\pm 0.17 $ & $1.06\pm 0.14 $ \\
Mrk 273 & $2.49\pm 0.42$ & $1.50\pm 0.20 $& $4.33\pm 0.33 $ \\
IRAS 15250+3609 & $6.99\pm 0.72 $ & $0.83\pm 0.09$ & $0.41\pm 0.04$ \\
IRAS 17208-0014 & $5.06\pm 0.47 $  & $1.18\pm 0.13 $ & $2.56\pm 0.20 $ \\
IRAS F08572+3915 & $1.31\pm 0.01 $  & $0.24\pm 0.04 $ & $0.51\pm 0.03$ \\
 \hline
\end{tabular}}
\caption{FIR line flux densities for the sample of ULIRGs used in this work. All of [O{\sc iii}]52 $\mu$m line observations are with SOFIA/FIFI-LS (Methods Section), while all of
  [O{\sc iii}]88 $\mu$m measurements are from the Herschel/PACS archive of SHINING survey of bright ULIRGs\cite{Herrera-Camus18b}.
  For [N{\sc iii}]57 $\mu$m, we also make use of the Herschel/PACS archive of SHINING survey, except for IRAS 15250+3609 which is a new measurement with SOFIA/FIFI-LS.}
\label{table:line_measurements}
\end{table*}

\bigskip
\bigskip

\begin{table*}[h]
\centering
\resizebox{1\columnwidth}{!}{%
\begin{tabular}{c c c c c c c} 
 \hline
Source ID & $z$ & $S_{60\mu m}\ (Jy)$& $\log(L_{IR}/L_\odot)$& $\log{M_*/M_\odot}$ & $\rm 12+\log(O/H)_{FIR}$ & $\rm 12+\log(O/H)_{Optical}$ \\ [0.5ex] 
 \hline\hline
IRAS 12112+0305& 0.0730 & 8.18& 12.48& 11.05 & $9.02^{+0.22}_{-0.16}$ & 8.75 \\
Mrk 273& 0.0378 &22.51&12.10& 10.84 & $9.13^{+0.09}_{-0.08}$& 8.77 \\
IRAS 15250+3609& 0.0552 &7.10&12.00& 10.67 & $9.03^{+0.09}_{-0.07}$ & 8.65 \\
IRAS 17208-0014& 0.0430 &34.79&12.68& 11.30 & $9.06^{+0.07}_{-0.06}$  & 8.94\\
IRAS F08572+3915 & 0.0584& 7.30&12.04 & 10.51 & $8.99^{+0.09}_{-0.08}$& 8.74 \\
 \hline
\end{tabular}}
\caption{The properties of the ULIRG sample used in this work. $z$: redshift, $S_{60\mu m}$: IRAS 60 $\mu$m flux density, $L_{IR}$: total 8-1000 $\mu$m IR luminosity, $M_*$: stellar mass, $\rm 12+\log(O/H)_{FIR}$: Oxygen abundance using [O{\sc iii}]52, 88 and [N{\sc iii}]57, $\rm 12+\log(O/H)_{Optical}$: Oxygen abundance using rest-frame recombination optical lines using the $R23$ or $O3N2$ estimators
  described in the Methods Section. The oxygen abundance measurements have a typical uncertainty of $0.1$ dex. }
\label{table:2}
\end{table*}

\newpage

\begin{addendum} 

\item[Acknowledgements]
Results in this paper are based on observations made with the NASA/DLR Stratospheric Observatory for Infrared Astronomy (SOFIA). SOFIA is jointly operated by the Universities Space Research Association, Inc. (USRA), under NASA contract NAS2-97001, and the Deutsches SOFIA Institut (DSI) under DLR contract 50-OK-0901 to the University of Stuttgart. Financial support for this work in part was also provided by NASA through award 80NSS20K0437. JLW acknowledges support from an STFC Ernest Rutherford Fellowship (ST/P004784/2).

\item[Author Contributions] NC and AC authored the draft version of this paper. NC measured line fluxes and FIR metallicities of the sample and conducted the analysis of this paper. AC, JM, HN and JLW were PI/co-I in the successful SOFIA proposal and performed the observations. All other coauthors contributed extensively in interpreting the results of this paper and provided extensive comments on this manuscript as part of an internal review process.

\item[Correspondence]

{Correspondence for materials should
be addressed to N.C. (nchartab@uci.edu) and A.C. (acooray@uci.edu)}

\item[Competing interests statement] The authors declare no competing
interests.

\end{addendum}

\clearpage

\begin{center}
{\bf \large Methods}
\end{center}
\vspace{-0.6cm}

Here we provide details of SOFIA/FIFI-LS observations combined with Herschel/PACS archival data of the FIR fine-structure lines of 5 ultra-luminous infrared galaxies (ULIRGs) with $S_{60\mu m} \geq 7$ Jy at $0.035 < z < 0.075$ (\S~\ref{sec:sample} and \S~\ref{sec:data}). We then present some details about FIR-based metallicity measurements (\S~\ref{sec:far-IR metallicity}) and stellar masses as well as optical-based metallicity measurements used in this work (\S~\ref{sec:Optical metallicity}). Throughout this paper, we adopt a Chabrier\cite{Chabrier03} initial mass function (IMF), and the concordance $\Lambda$CDM cosmology\cite{Komatsu11} with $\Omega_{\rm m}=0.3$, $\Omega_\Lambda=0.7$, and $H_0$ = 70~km~s$^{-1}$~Mpc$^{-1}$.

\section{The sample of Ultra-luminous Infrared Galaxies}
\label{sec:sample}
The Herschel ULIRG Survey (HERUS\cite{Farrah13}) and Survey with Herschel of the ISM in Nearby INfrared Galaxies (SHINING\cite{Sturm11,Herrera-Camus18a,Herrera-Camus18b}, overlapping with ULIRGs in the Great Observatories All-Sky LIRG Survey (GOALS) sample\cite{Diaz-Santos17}), assembled Herschel/PACS and SPIRE observations of 43 ULIRGs at $z < 0.27$ drawn from the IRAS PSC-z survey\cite{Saunders00}, with a 60 $\rm \mu$m flux density greater than $\sim 1.7 \rm Jy$, and made detailed diagnostics of the FIR fine-structure lines possible\cite{Farrah13,Herrera-Camus18a,Pereira-Santaella17}. All objects have been spectroscopically confirmed with high-resolution optical observations and observed with Spitzer IRS for mid-IR lines and PAH features\cite{Armus07,Farrah07,Desai07}.

Although HERUS, SHINING, and GOALS surveys already represent the best-explored samples so far with some FIR lines observed such as [O{\sc i}]63 $\mu$m and [C{\sc ii}]158 $\mu$m, none of the objects have the complete set of lines simultaneously observed, missing critical set of lines [O{\sc iii}]52 $\mu$m, [N{\sc iii}]57 $\mu$m and [O{\sc iii}]88 $\mu$m for the study on metal abundances reported in the main Letter. These three emission lines are crucial since it has been shown that the flux line ratio of (2.2$\times$[O{\sc iii}]88+[O{\sc iii}]52)/[N{\sc iii}]57 provides a powerful tool for gas-phase metallicity of galaxies, especially for highly obscured star-forming galaxies\cite{Pereira-Santaella17}. 

Among these three lines, [O{\sc iii}]52 $\mu$m line has been challenging to measure for local galaxies and ULIRGs. In particular, Herschel Space Observatory\cite{Pilbratt10} provided far-IR spectroscopy with unprecedented sensitivity and spatial resolution. However, the [O{\sc iii}]52 $\rm \mu$m line lies outside the 55 to 210 $\rm \mu$m spectral range of PACS\cite{Poglitsch10} for $z < 0.06$ objects. Herschel completed science operations in May 2013. The SOFIA observatory\cite{Temi18} is now in its full science operations and provides the spectroscopic coverage over the necessary wavelength range for measuring the full far-IR metallicity diagnostic.

We initially selected 11 ULIRGs from HERUS, SHINING, and GOALS surveys but reduced the sample to 5 ULIRGs that have [O{\sc iii}]52 $\mu$m line uncontaminated by the atmospheric transmission window accessible with SOFIA/FIFI-LS. In the main Letter, we present SOFIA/FIFI-LS observation of these 5 local ULIRGs with $S_{60\mu m} \geq 7$ Jy at $0.035 < z < 0.075$ which is conducted to obtain missing lines from the full set of [O{\sc iii}]52 $\mu$m, [O{\sc iii}]88 $\mu$m, and [N{\sc iii}]57 $\mu$m lines. These lines are necessary for measuring robust estimates of gas-phase metallicity that are almost independent of the ionizing source and ionization parameter\cite{Pereira-Santaella17}.

\section{SOFIA/FIFI-LS Data}
\label{sec:data}
FIFI-LS\cite{Fischer18} is a far-IR IFU spectroscopy instrument in the mid-IR wavelengths onboard SOFIA, with the design and capabilities similar to those of the Herschel/PACS instrument but with a shorter spectral wavelength cutoff than PACS. In the 51–125 $\rm \mu$m blue channel, it has 5 $\times$ 5 spatial pixels (spaxels) of size $6$ arcseconds per pixel. We obtained far-IR spectroscopic observations of all our sources from February 2020 to June 2021 during Cycles 8 and 9 of SOFIA operations using the FIFI-LS instrument (PID:08\_0095). Our main target line was [O{\sc iii}]52 $\mu$m line as it is not covered by Herschel/PACS data for any of our local sources. However, we also observed [N{\sc iii}]57 $\mu$m line of IRAS 15250+3609 as it is not available with Herschel/PACS in the SHINING archive.
All the SOFIA data have gone through the Level 4 pipeline reduction at the SOFIA Science Center. The analysis includes the conversion of raw data to astronomical observations, including astrometry and flux calibrations, as well as corrections for the instrument response and atmospheric transmission.

The Level-4 spectral data cube for each ULRIG was further analyzed for scientific measurements using the Python GUI software SOSPEX\cite{Fadda18}. We extract the 1D spectra by integrating the spectral images over an area corresponding to the existing Herschel/PACS observations (Fig. \ref{fig:spectra}).
We fit a single Gaussian function to continuum-subtracted spectra centered at the known peak emission of galaxies based on their confirmed redshifts to measure line flux densities and their corresponding uncertainties. We perturb the spectra using their error distributions and estimate the line flux ratio for a hundred trials. The average and standard deviation of hundred trials are adopted as the line flux density and its uncertainty, respectively. The line flux maps and continuum-subtracted spectra for all the data are presented in Figure \ref{fig:spectra}. We also plot an ellipse in each line map that shows the region from which the spectrum is extracted. 

\subsection{Ancillary Data}
Herschel/PACS\cite{Poglitsch10} provided the most extensive archive of spectral observations of [N{\sc iii}] and [O{\sc iii}] lines. [N{\sc iii}]57 $\mu$m and [O{\sc iii}]88 $\mu$m spectra are taken by Herschel/PACS for all our objects except for IRAS 15250+3609. A detailed description of the data reduction and flux measurements of [N{\sc iii}]57 $\mu$m and [O{\sc iii}]88 $\mu$m lines can be found in the SHINING data release paper\cite{Herrera-Camus18a}. [N{\sc iii}]57 $\mu$m line flux of IRAS 15250+3609 is measured from SOFIA/FIFI-LS observations. All the line fluxes for our sample are given in Table \ref{table:line_measurements} of the main Letter. 

\begin{figure*}[ht]
    \centering
    \subfloat{{\includegraphics[width=.47\linewidth,clip=True, trim=0cm 0cm 0cm 0cm]{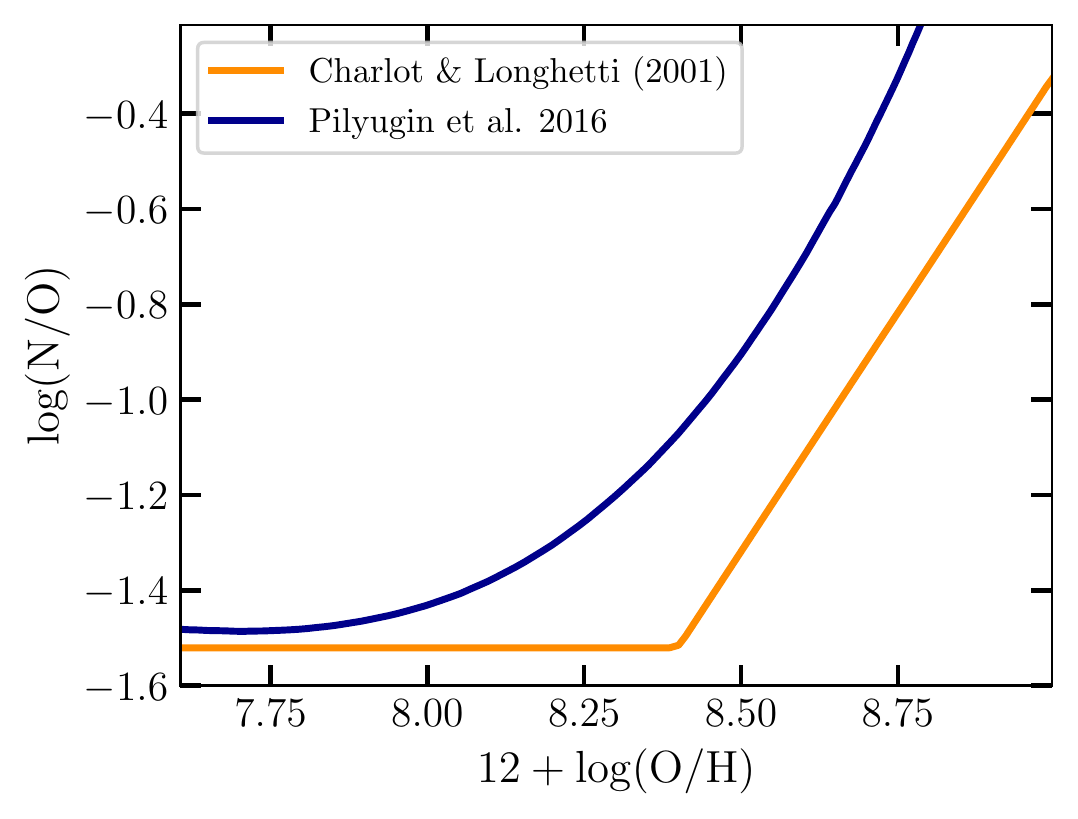} }}%
    \qquad
    \subfloat{{\includegraphics[width=.45\linewidth, trim=0cm 0cm 0cm 0cm]{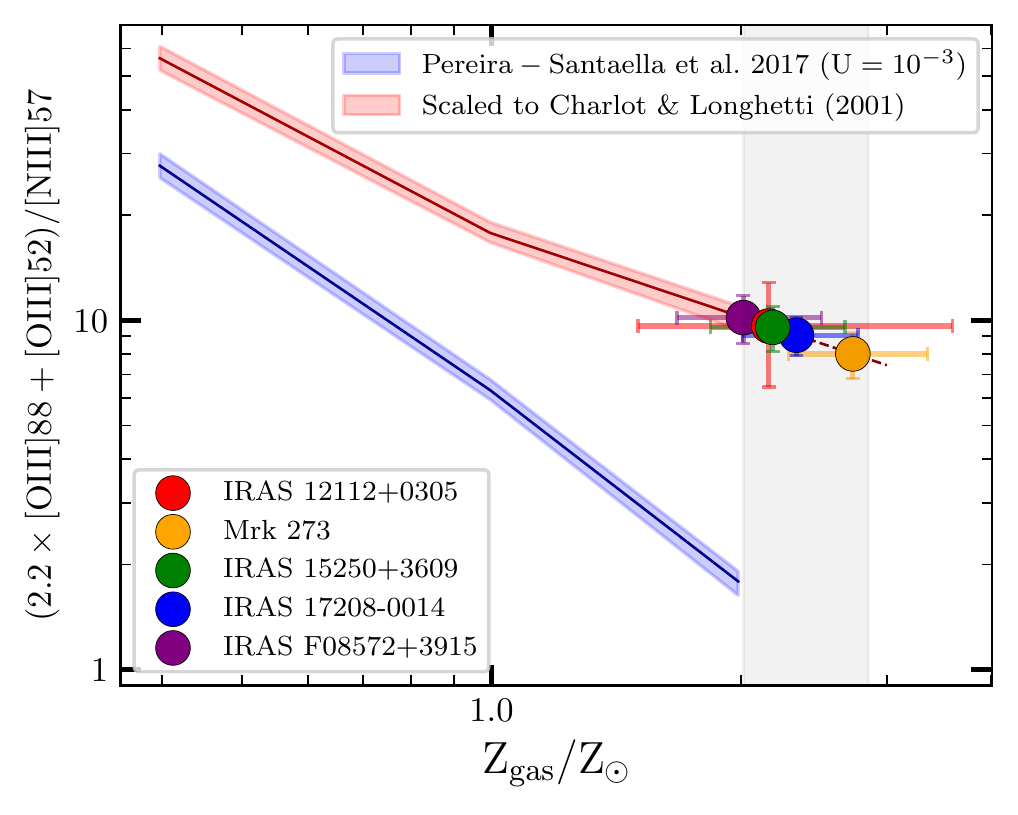} }}%
    \caption{{\it Left:} $\rm N/O$ as a function of $\rm O/H$. The blue line is a third-order polynomial fit to the observed $\rm N/O$-$\rm O/H$ relation\cite{Pilyugin16}, which is employed in FIR calibration. The calibration for optical metallicity is based on a relation\cite{Charlot01} which is shown with an orange line. The discrepancy between the two models directly affects the metallicity measurements and one needs to correct this systematic offset resulting from different assumptions. {\it Right:} The blue line shows the FIR metallicity calibration employing observed $\rm N/O$-$\rm O/H$ relation (blue line in the left panel), while the red line shows the scaled relation which is corrected for the systematic difference between two $\rm N/O$-$\rm O/H$ models. We use the scaled relation to measure the metallicities to maintain comparable results with optical measurements. For the solar metallicity, we adopt $\rm 12 + \log(O/H)_\odot = 8.69$\cite{Asplund09}.}%
    \label{fig:calibration}
\end{figure*}

\section{Far-IR metallicity}
\label{sec:far-IR metallicity}
A recent study\cite{Herrera-Camus18a} revisited the mass-metallicity relation of the SHINING (U)LIRGs with metallicities constrained from the [O{\sc iii}]88/[N{\sc iii}]57 ratio. They showed that (U)LIRGs tend to lie below the Mass-Metallicity Relation (MZR) of star-forming galaxies, but the offset is smaller than previously thought from studies based on optical-based metallicities. In this paper, for the first time, we use a robust tracer, (2.2$\times$[O{\sc iii}]88+[O{\sc iii}]52)/[N{\sc iii}]57, to measure far-IR metallicities for our ULIRG sample and draw definite conclusions on whether the optical-based metallicities are underestimated thus cannot represent the ULIRG sample. We convert line ratios, (2.2$\times$[O{\sc iii}]88+[O{\sc iii}]52)/[N{\sc iii}]57, to gas-phase metallicity using a calibration introduced by Ref.\cite{Pereira-Santaella17}, considering the ionization parameter value of $\rm U=10^{-3}$. For our sample with $\rm \Sigma_{FIR}\sim 10^{11}-10^{12}$ $\rm L_\odot/Kpc^2$, an ionization parameter of $\rm U\sim 10^{-2.7}-10^{-3.5}$ is estimated from observations\cite{Herrera-Camus18b}. This variation has a negligible effect on our metallicity measurements ($<0.05$ dex).

To directly compare our results with metal abundances measured for the same ULIRGs with fine-structure lines at optical wavelengths\cite{tre04}, we scale our gas-phase metallicity measurements to take into account the differences in the FIR metallicity calibration\cite{Pereira-Santaella17} and optical metallicity calibration\cite{tre04}. The FIR-based metallicity calibration\cite{Pereira-Santaella17} utilizes an observational relation for $\rm N/O$ vs $\rm O/H$ while optical metallicity calibration\cite{tre04} employs a theoretical model\cite{Charlot01} to convert from N/O ratio to O/H estimates. These two relationships, one observational and one from a model, are shown in Figure \ref{fig:calibration}, suggesting that assumptions regarding the behavior of $\rm N/O$ as a function of $\rm O/H$ directly affect the metallicity measurements. For FIR metallicity, we scale the calibration\cite{Pereira-Santaella17} to allow a direct comparison with existing optical measurements and the MZR relation. In effect, we have converted our abundance measurements, first calibrated with the observed relation\cite{Pereira-Santaella17} to that of the model used in SDSS work\cite{tre04} (right panel in Figure \ref{fig:calibration}). 

\begin{figure}[h]
    \centering
    \includegraphics{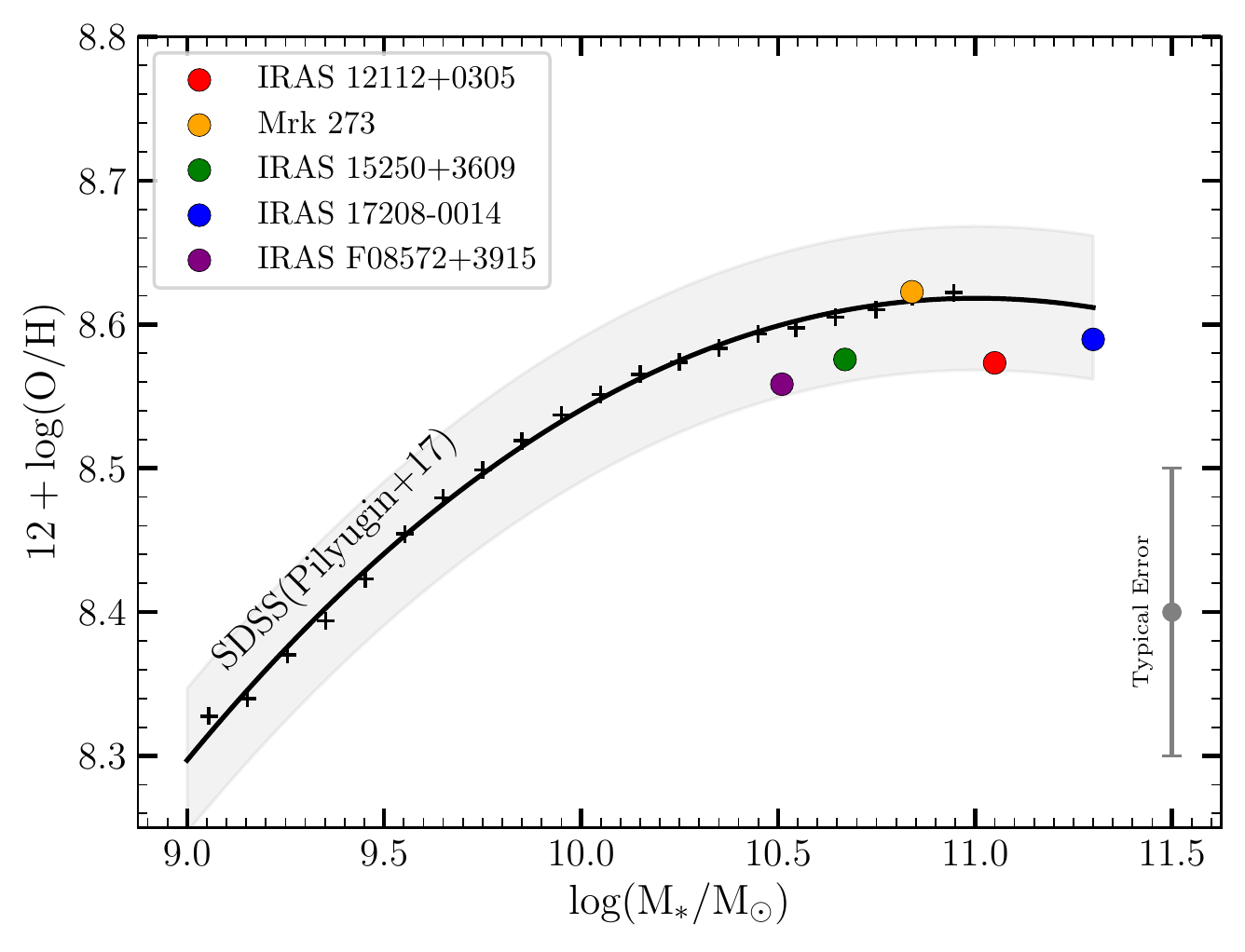} 
    \caption{Similar to Figure \ref{fig:MZR} of the main Letter, but FIR metallicities are derived from the original calibration without scaling values to the theoretical model of $\rm N/O-\rm O/H$ relation. The MZR of SDSS star-forming galaxies is shown from calibration with a consistent assumption of $\rm N/O-\rm O/H$ relation as FIR metallicity. The plus signs showing the MZR of star-forming SDSS galaxies at $z=0.05$ are taken from literature\cite{Pilyugin17}. The black line is a third-order polynomial fit to SDSS MZR data points. }
    \label{fig:MZR_PG}
\end{figure}

Further, we explore whether re-calibrating the metallicities of the FIR can affect our results. We use the original FIR metallicity calibration\cite{Pereira-Santaella17} without scaling it to the theoretical model of $\rm N/O-\rm O/H$ relation\cite{Charlot01}. Figure \ref{fig:MZR_PG} shows the comparison of the FIR metallicities derived from original calibration and MZR of SDSS star-forming galaxies\cite{Pilyugin17}, which are calculated with the consistent assumption of $\rm N/O-\rm O/H$ relation as FIR metallicity. We find that ULIRGs lie on the MZR of star-forming galaxies without any evidence of significant under-abundance, confirming our findings in the main Letter.

\section{Stellar masses and optical metallicities}
\label{sec:Optical metallicity}
In Figures \ref{fig:MZR} and \ref{fig:SDSS} of the main Letter, the stellar mass of IRAS 12112+0305, Mrk 273, IRAS 17208-0014, and IRAS F08572+3915 are adopted from the measurements\cite{U12} based on the SED fitting of the UV-NIR part of the spectrum. Stellar masses of 15250+3609 and IRAS 17208-0014 are derived from SED fitting over NIR-Radio and UV-Radio part of the spectrum, respectively\cite{daCunha10,Vega08}. Our sample consists of galaxies with $\rm 10^{10.5}M_\odot<M_*<10^{11.3}M_\odot$, which are located in the massive end of the MZR.

The optical metallicities of IRAS 15250+3609 and IRAS F08572+3915 are taken from literature and converted to the desired calibration\cite{tre04} if they are computed using different calibration\cite{Monreal-Ibero07,Rupke08}. For the other three sources, we use O3N2=([O{\sc iii}]5007/H$\beta$)/([N{\sc ii}]6584/H$\alpha)$ indicators to estimate the oxygen abundances as all these optical lines are available for IRAS 17208-0014, Mrk 273 and IRAS 12112+0305\cite{Hou09}. All oxygen abundance estimates are converted to the desired calibration\cite{tre04}. We find that based on optical metallicities, ULIRGs have a significant offset ($\sim 0.3$ dex) from the MZR, which is consistent with previous studies\cite{Rupke08}. However, when we use FIR-based gas-phase metallicities, such offset disappears. Previous studies found that luminous optically-selected mergers are under-abundant, $\lesssim 0.1$ dex\cite{Kewley06}. Our results imply that ULIRGs follow similar luminous star-forming galaxies and unusual under-abundance ($\sim 0.3$ dex) found in optical metallicities is a result of heavily obscured metal-rich gas which has a negligible effect when using the FIR line diagnostics.       

Shown in Figure \ref{fig:SDSS} of the main Letter, we draw a control sample of star-forming galaxies from the MPA/JHU Value-Added Galaxy Catalog of SDSS-DR7\cite{Kauffmann03,tre04,Brinchmann04} with a similar stellar mass ($\rm 10^{10.5}M_\odot\lesssim M_*\lesssim 10^{11.5}M_\odot$) and star formation rate ($\rm SFR\sim 10^{2.5}\ M_\odot/year $) to our ULIRG sample, excluding extremely dusty galaxies ($\rm A_v<2.4$). We find that these less dusty galaxies have an average optical metallicity of 9.1, while the average optical metallicity of our sample is 8.8. One might conclude that the $0.3$ dex offset implies that the ULIRG sample does not follow the stellar mass-SFR-metallicity relation, known as fundamental metallicity relation\cite{Mannucci10}. However, optical lines are susceptible to extinction and thus, FIR metallicity calibration provides a reliable measurement for gas-phase metallicity of heavily dust-obscured systems. We find that FIR metallicity measurements of ULIRGs with an average of 9.05 are consistent with the average metallicity of galaxies with similar stellar mass and star formation rate. Therefore, our ULIRG sample follows the fundamental metallicity relation and optical lines are not representative of the metallicity of heavily dust-obscured galaxies.

\begin{figure*}[h]
    \centering
    \includegraphics[width=\linewidth,clip=True, trim=0cm 0cm 0cm 0cm]{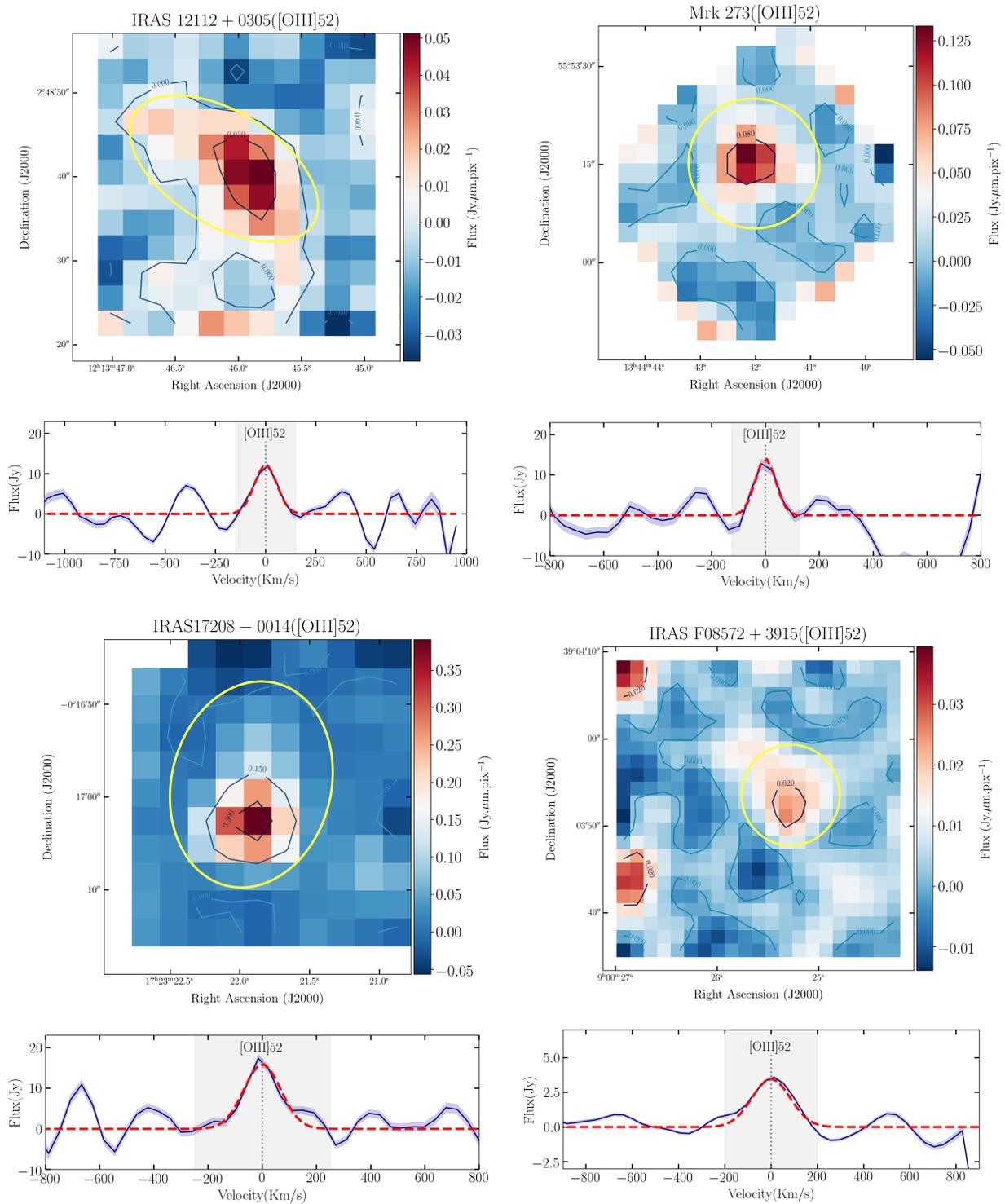}
    \caption{Line maps and spectra of our sample targeted by SOFIA/FIFI-LS. }
    \label{fig:spectra}

\end{figure*}

\begin{figure*}
\ContinuedFloat
    \centering
    \includegraphics[width=\linewidth,clip=True, trim=0cm 0cm 0cm 0cm]{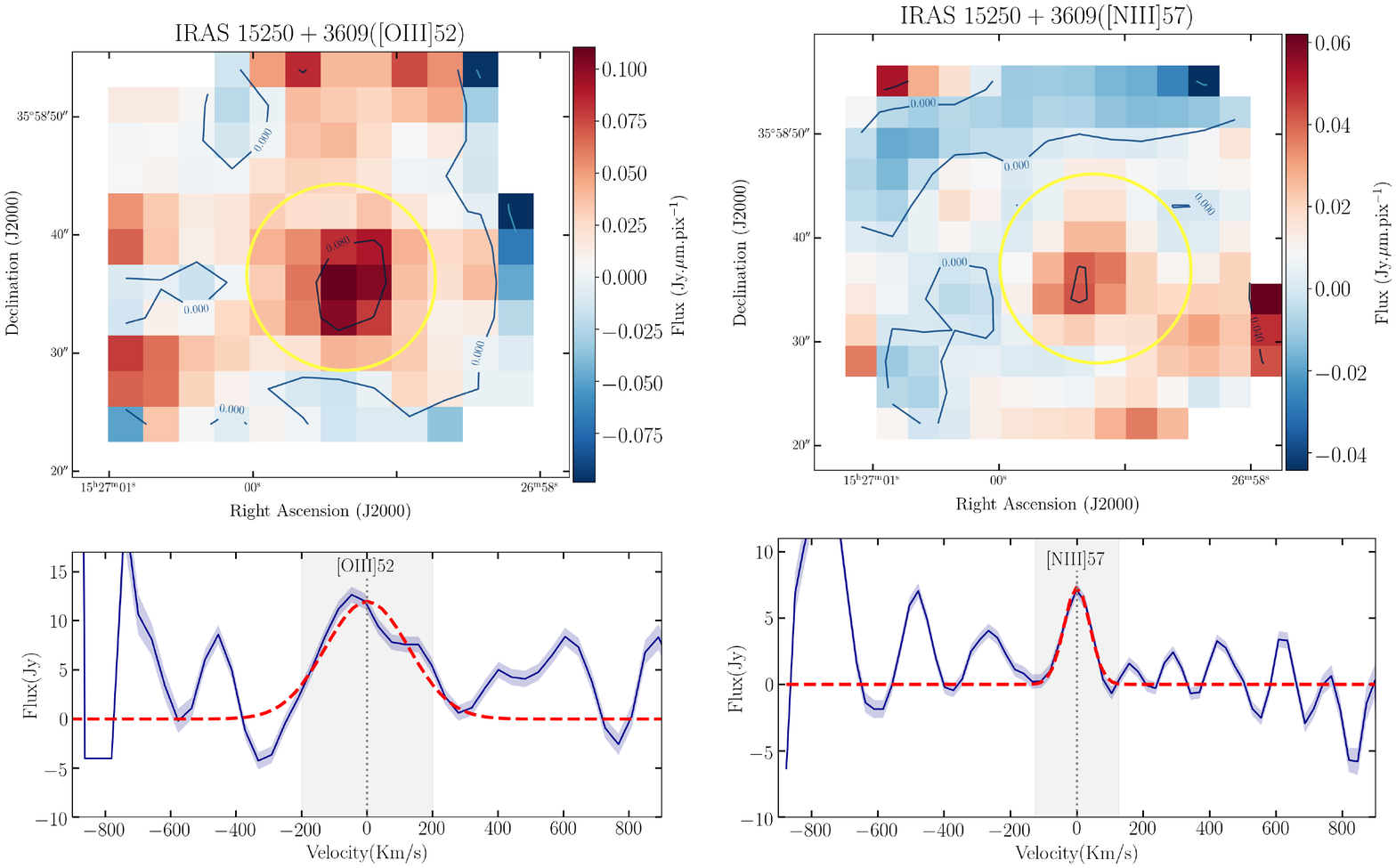}
    \caption{Continued: Each panel shows the moment 0 map of the spectral line with a yellow ellipse that demonstrates the region over which the spectrum is measured. In the bottom sub-panels, the extracted spectrum is shown along with the best-fit Gaussian function. The blue shaded region around the spectrum corresponds to the 1$\sigma$ uncertainty of the spectrum. The gray shaded regions on each spectrum show the range of velocity in which the 0$^{th}$ moment line maps are calculated.}
\end{figure*}

\clearpage

\begin{center}
\bf{\large Methods References}
\end{center}

\section{Data Availability Statement}
Data supporting this study is publicly available or will be available by June 2022 through the \href{https://irsa.ipac.caltech.edu/applications/sofia/}{NASA/IPAC Infrared Science Archive}, under Plan ID of 08\_0095. Currently, SOFIA/FIFI-LS observations are publicly available for IRAS F08572+3915 and Mrk273 and will be made available for the remainder of the sources by June 2022.

\end{document}